# Controlling Optical Quantum Interference with Plasmonic Metasurfaces

Lawrence M. Qin,[1,2,#] Sky Semone,[1,2,#] Matthew J. Brandsema,[2] and Christos Argyropoulos[1,*]

[1]Department of Electrical Engineering, Pennsylvania State University, University Park, PA, 16803 USA
[2]Applied Research Lab, Pennsylvania State University, University Park, PA, 16803 USA

[#]These authors contributed equally to this work.
*cfa5361@psu.edu

## Abstract

The ability to efficiently control optical quantum interference by the same device is a challenging task that promises to enable key operations in quantum information processing. Here, we tackle this problem by presenting a composite quantum plasmonic metasurface operating as an ultrathin non-unitary beamsplitter capable of achieving both single photon bosonic bunching and fermionic anti-bunching interference just by varying the incident photons frequency or changing the design dimensions. The presented metasurface combines a plasmonic grating corrugated by periodic nanoslits that can achieve extraordinary optical transmission with additional nanostripes on top forming nanocavities characterized by localized gap plasmon resonances. The photon-photon interactions are analyzed based on calculating the quantum coalescence metric which accurately portrays the obtained bunching or anti-bunching quantum interference. The presented quantum interference operation is applicable to both fundamental non-classical light interactions and in the design of more applied quantum optical logic gates used in quantum computing and communications.

## Introduction

The Hong-Ou-Mandel (HOM) effect is a fundamental phenomenon in quantum engineering that demonstrates the wave-like interference of two individual photons [1,2]. It occurs when two identical single photons enter a 50/50 lossless beam splitter, one through each input port. When two indistinguishable single photons arrive at such a beam splitter at the exact same time, destructive interference occurs, which can eliminate the possibility of the photons exiting through different ports. As a result, the photons "bunch" together and always exit the beam splitter through the same output port, also known as photon bunching. By measuring the coincidence counts (i.e., how often both single photon detectors click at the same time) while varying the arrival time of one photon, a sharp drop is observed in coincidences when the photons are perfectly synchronized. This response cannot occur from conventional light sources characterized by classical electromagnetic theory and only quantum single photon sources described by quantum electrodynamics can realize such an effect.

Conventional unitary beamsplitters are lossless optical components of paramount importance and widely used by engineers working in quantum optics and photonics. These ideal devices are extensively utilized to demonstrate one of the most important properties of bosonic quantum single photon-photon interactions, named photon bunching. While lossless 50:50 beamsplitters can be used to demonstrate the basic principles of the HOM effect and associated bosonic behavior, the introduction of non-unitary beamsplitters [3] with radiative/non-radiative losses can realize additional degrees of freedom to quantum single photon-photon interactions leading to fermionic behavior usually verified by photon anti-bunching.

Recently, quantum metasurfaces have emerged as a topic of substantial interest in mimicking, destroying, or tuning the HOM effect, effectively operating as new compact nanoscale designs of non-unitary beamsplitters [4–6]. More specifically, with the advancements in plasmonic and dielectric metasurfaces, several structures have been designed that enrich the potential of beamsplitters in controlling the HOM effect [7–12], such as causing the incident photons to forcibly separate out of the metasurface ports, imitating the behavior of fermionic particles and realizing an anti-bunching response [8,11,12]. These two opposite optical quantum interference behaviors lead to the quantum metasurface coalescing and anti-coalescing, which will be the main terminology used here to describe the bunching bosonic behavior and anti-bunching fermionic response, respectively [10,11,13].

Another recent approach allowed a metasurface to demonstrate either coalescing or anti-coalescing behaviors based on quantum graphs encoded via Jones matrices, where multiphoton bunching and anti-bunching were controlled by designing the fabricated matrices to have a certain degree of overlap across different diffraction orders [14]. Moreover, the Pancharatnam-Berry phase profile was realized with a quantum metasurface with the goal to dictate coalescence and anti-coalescence behavior by engineering the orthogonality of the metasurface's output ports [15]. While most previous metasurface designs were dielectric, metallic gap-plasmon nanoantenna arrays forming plasmonic metasurfaces were also used to engineer the anisotropic quantum vacuum achieving remote quantum interference and control in coalescence response through changes of the metasurface's orientation-dependent cross-damping [16].

For improved tunability that allows the metasurface to change coalescence in real time, researchers explored creating non-unitary lossy metasurfaces with anisotropic phase responses that allow such tuning through mechanical rotation [8]. This response alters the phase between the reflected and transmitted waves continuously tuning the photon-photon interactions from partially bunched to moderately anti-bunched states. Another relevant work performed similar optical quantum interference but with a unitary metasurface that decouples the horizontal and vertical components of the incident photon polarizations by having an anisotropic polarization response [7]. That allowed the polarized components of the transmitted and reflected waves to efficiently interfere with each other realizing modified HOM effect interference. Finally, the single photon-photon interference was modulated by physically rotating tunable q-plates which alter the internal geometric phases of the reflected and transmitted waves, causing an alternating optical quantum interference output varying between coalescing and anti-coalescing single photon streams along a sinusoidal curve [10].

Alternative approaches to tuning the quantum interference utilized controlled losses, specifically Coherent Perfect Absorption (CPA), to create metasurfaces that can achieve the anti-bunching fermionic effect. Initially, it was established that CPA works on single photons [9] and subsequent works demonstrated that losses could force a boson to undergo fermionic behavior based on lossy beam splitters [11] and CPA configurations [12] , leading to anti-coalescence. In particular, by placing the absorber at the anti-node of standing waves within the material, it can deterministically absorb specific symmetric or anti-symmetric wavefunctions, forcing identical bosonic photons into disparate ports to mimic fermionic anti-bunching behavior [9,11,12]. In the CPA case [12], the introduced losses create an altered phase response, where anti-symmetric waves are forced to act fermionically while symmetric waves act as bosons.

Tunable quantum two-photon interference was also achieved based on the integration of phase-change materials, such as Germanium Telluride (GeTe), into planar metasurface geometries [13]. This effect occurred by triggering a thermally driven crystallographic transition between amorphous and crystalline GeTe phases that can dynamically tune the intrinsic loss, hence, non-mechanically switching the quantum interference output from anti-coalescence to coalescence. As an alternative approach, liquid crystal metagratings were utilized whose birefringence is voltage controlled to create an electrically tunable beamsplitter [17]. By varying the voltage in such devices, the transmittance and reflectance were tuned due to the splitting ratio of the liquid crystal material enabling it to undergo controllable quantum bunching interference.

The emerging metasurface technology is proposed to revolutionize quantum optics leading to highly scalable, low-decoherence quantum information infrastructures by replacing bulky passive linear optical components with ultra compact flat optics. Specifically, unitary and tunable designs are intended to realize highly parallel quantum computing logic gates and rapid, single-shot quantum state tomography devices [8,13]. Non-unitary and CPA based designs can function as high dimensional entanglement filters to purify and protect robust quantum communication networks from environmental decoherence [11,18].

While all the metasurfaces described before have incredible potential, those that aim to tune the optical quantum coalescing response usually require some type of outside bias to change their coalescence behavior, mostly through mechanical [7,8,10] or electrically driven [17] external stimulation. In quantum communication and computing paradigms this external bias would put significant space and manufacturing burdens on the area surrounding the metasurface, which would create deficiencies in modern photonic devices where its dense circuitry and non-reliance on temperature-controlled environments make it stand out from conventional electronic counterparts.

To tackle this limitation, in our current work, we propose an ultrathin, passive quantum plasmonic composite metasurface with schematic seen in Fig. 1 that achieves perfectly varying coalescence and anti-coalescence response while eliminating the need of external bias or another localized electronic circuitry. Rather than tuning the optical properties of the material via external heat or mechanical stress, our composite asymmetric metasurface design dictates the two-photon interference state strictly via the onset frequency of the incident photons. By combining relatively broadband extraordinary optical transmission (EOT) response with a highly localized, narrowband magnetic gap-plasmon resonance, the device can transition between fermionic anti-coalescence and bosonic coalescence by using incident photons of slightly different frequencies.

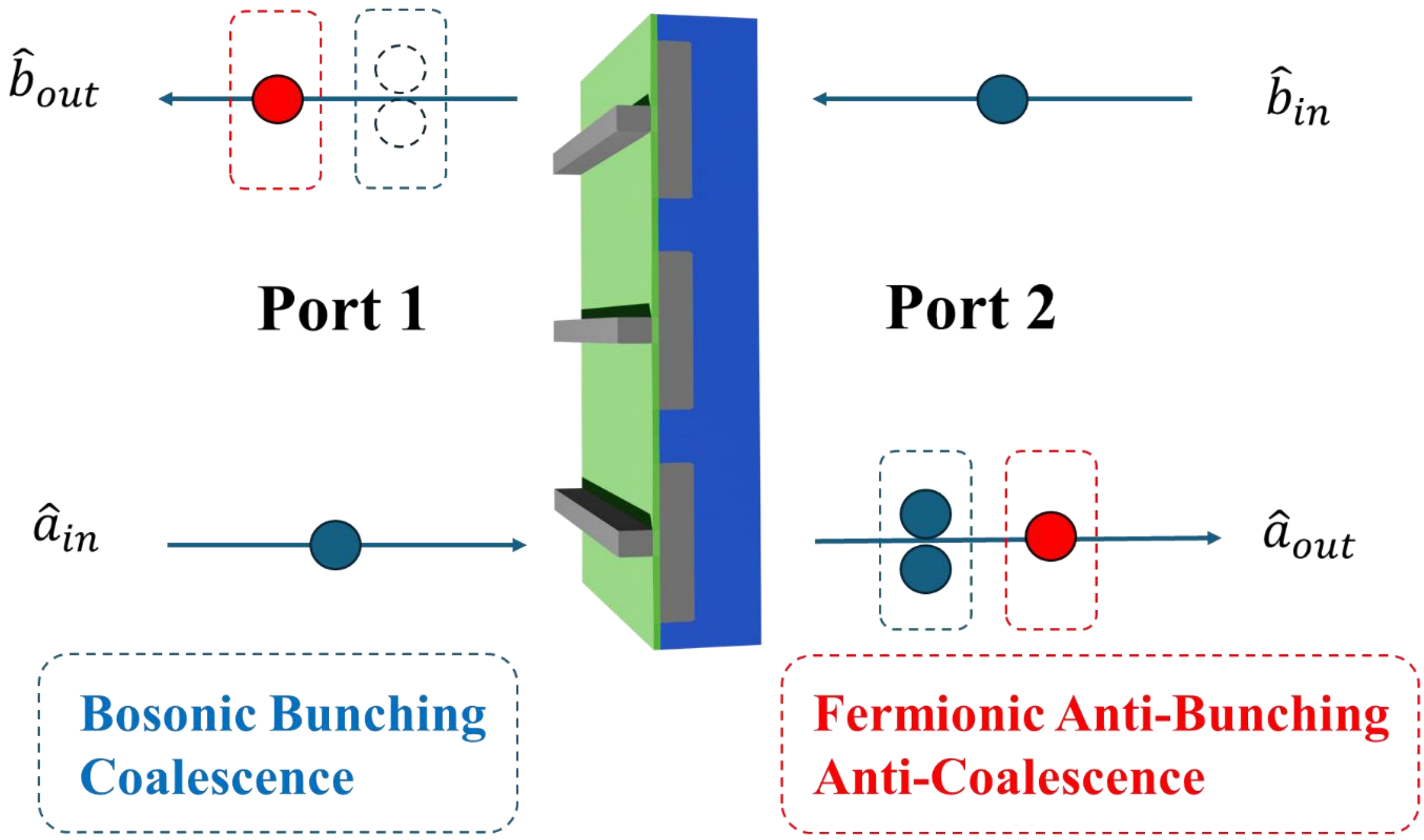

**Fig. 1: Quantum plasmonic metasurface.**

Schematic of the proposed quantum metasurface with two identical photons incident from two defined ports located left and right of the metasurface. The optical quantum interference induced by the metasurface forces the photons to act either bosonically (blue box) or fermionically (red box) at different operation frequencies.

## Results

### Theory

First, we calculate the joint probability of single photons exiting from either port 1, port 2 or both ports, as seen in Fig. 1, with the goal to quantitatively characterize the nonclassical two-photon interference. In the general case of a spatially asymmetric but reciprocal device, the forward ($r_{11}$) and backward ($r_{22}$) side reflection coefficients are different, but the transmissions ($t_{12}, t_{21}$) from both sides are equal and can be combined into a single transmission coefficient ($t_{12} = t_{21} = t$). Reflectance, transmittance, and absorption can be computed by these coefficients for the forward and backward directions as $R_{11} = |r_{11}|^2$, $R_{22} = |r_{22}|^2$, $T = |t|^2$, $A_{12} = 1 - |r_{11}|^2 - |t|^2$, $A_{21} = 1 - |r_{22}|^2 - |t|^2$. Lossless beamsplitter theory assumes a unitary condition where $|t|^2 + |r_{11}||r_{22}| = 1$. Hence, the relationship between the input annihilation operators of ports 1 and 2 $(\hat{a}_{in}, \hat{b}_{in})$ and their corresponding output operators $(\hat{a}_{out}, \hat{b}_{out})$, all of them showcased in Fig. 1, strictly rely on the transmission and reflection coefficients. However, realistic devices are dispersive and lossy, and we need to loosen the transmission and reflection restrictions ($|t|^2 + |r_{11}||r_{22}| < 1$) by introducing Langevin noise operators to account for the resulted dissipated-driven quantum interactions [19]. More details about the asymmetric beamsplitter quantum theory are presented in Supplementary Information section 1.

It should be noted that since the photons will be coming from the same source, they are assumed to be indistinguishable. Thus, a superposition forms and the quantum probabilities to achieve bosonic bunching ($P(2,0)_q$ and $P(0,2)_q$), meaning two photons are exiting bunched together either in port 1 or 2, or fermionic anti-bunching ($P(1,1)_q$), i.e., two photons are exiting separately in port 1 and 2, responses (see Fig. 1 for more details) are given below [7,19,20]:

$$P(2,0)_q = 2|t|^2|r_{11}|^2, \tag{1}$$

$$P(0,2)_q = 2|t|^2|r_{22}|^2, \tag{2}$$

$$P(1,1)_q = |t|^4 + |r_{11}|^2|\, r_{22}|^2 + 2|t|^2|r_{11}||r_{22}|\cos(\theta_{\mathrm{rt}}\,). \tag{3}$$

Note that in the case of fermionic anti-bunching ($P(1,1)_q$), the photons interfere with each other based on the phase difference between the combined reflection ($|r_{11}||r_{22}|$) and transmission ($|t|^2$) coefficients of the system given by: $\theta_{rt} = \angle_{r_{11}} + \angle_{r_{22}} - 2\angle_t$. This is consistent with the Hong Ou Mandel effect as having 180 degrees phase shift created by a standard lossless beamsplitter cause $P(1,1)_q = (|t|^2 - |r_{11}||r_{22}|)^2$, where this fermionic response probability cancels when the amplitude condition $|t|^2 = |r_{11}||r_{22}|$ is satisfied, leaving only the bosonic probabilities ($P(2,0)_q, P(0,2)_q$) to not cancel out. As a result, when indistinguishable photons are incident on the lossy beamsplitter from the input ports $\left(\hat{a}_{in}, \hat{b}_{in}\right)$, they will only leave the beamsplitter "bunched together" as photon pairs also highlighted in blue in Fig. 1. However, if the phase shift is 0 degrees, the fermionic response probability becomes $P(1,1)_q = (|t|^2 + |r_{11}||r_{22}|)^2$ and cannot be equal to zero. Thus, when these photons interact with the lossy beamsplitter, they are strongly pushed to act fermionically, as separate photons, instead of bunched together, which is demonstrated in Fig. 1 red dashed boxes. The fermionic probability for classical distinguishable particles (a.k.a., classical probability) is equal to its quantum counterpart given by Eq. (3) but when the complex transmission and reflection coefficients are orthogonal and have a phase difference of $\theta_{rt} = 90°$ [3,8,20]:

$$P(1,1)_c = |t|^4 + |r_{11}|^2|r_{22}|^2. \tag{4}$$

Based on these probabilities that define anti-bunching and bunching optical quantum interference, we can compute the quantum coalescence. This quantum metric measures photon bunching detected by the HOM effect in unitary (lossless) beam splitters, while anti-coalescence is the measure of the opposite anti-bunching effect that forces quantum light to exit from different ports. To measure the coalescence metric (*C*), we need to compare the fermionic quantum probability of a lossy beamsplitter to the classical probability where the particles are distinguishable and do not constructively or destructively interfere leading to the formula [7,13]:

$$C = \frac{P(1,1)_q}{P(1,1)_c} = 1 + \frac{2|t|^2|r_{11}||r_{22}|\cos(\theta_{rt})}{|t|^4 + |r_{11}|^2|r_{22}|^2}. \quad (5)$$

This clearly shows that there is a direct impact of the incident classical wave phase and amplitude on the coalescence of the system. The amplitude of the reflected and transmitted waves determines the probability of the effect, while the phase dictates the nature of the quantum interaction which can either be bosonic or fermionic. For example, in a standard unitary beamsplitter, the phase of the reflected and transmitted waves are 180 degrees offset and their amplitudes are equal. In a lossy beamsplitter system, as $\theta_{rt}$ approaches 180 degrees, the destructive interference of the single photon probabilities increases until it reaches the threshold where the system shows standard bosonic bunching behavior (*C=0*). However, when the metasurface, or any lossy beamsplitter device, approaches phase differences of 0 degrees, these two paths undergo constructive interference instead, causing the system to start forcing photon pairs to separate; meaning photons exiting from separate ports, i.e., undergoing anti-coalescence (*C=2*).

While the phase of the incident waves changes the nature of the interaction, the amplitudes of the reflected and transmitted waves determine the strength of this interaction. The closer their amplitudes are to the relation $|t|^2 = |r_{11}||r_{22}|$, the stronger the interference is. This is numerically demonstrated in Fig. 2, where we fix $|t| = 0.5$ and obtain a clear trend shown by the dashed white lines when $\theta_{rt} = 0°$ and $\theta_{rt} = 180°$ that causes C to approach 2 and 0, respectively, in the case of interference approaching the aforementioned amplitudes relation. Note that when one of the reflections approaches zero (no reflection) or one (only reflection), it cancels the fraction term of C, which causes it to approach one, thus the photons start to act like classical distinguishable particles. This is also valid when $\theta_{rt} = 90°$, which causes the complete cancellation of the fraction term where all reflection amplitude values lead to *C=1*, as also clearly depicted in Fig. 2. Finally, it should be noted that the reflection coefficient amplitudes plotted as x- and y-axes in Fig. 2 are limited to approximately 0.87 due to energy conservation law when $|t| = 0.5$.

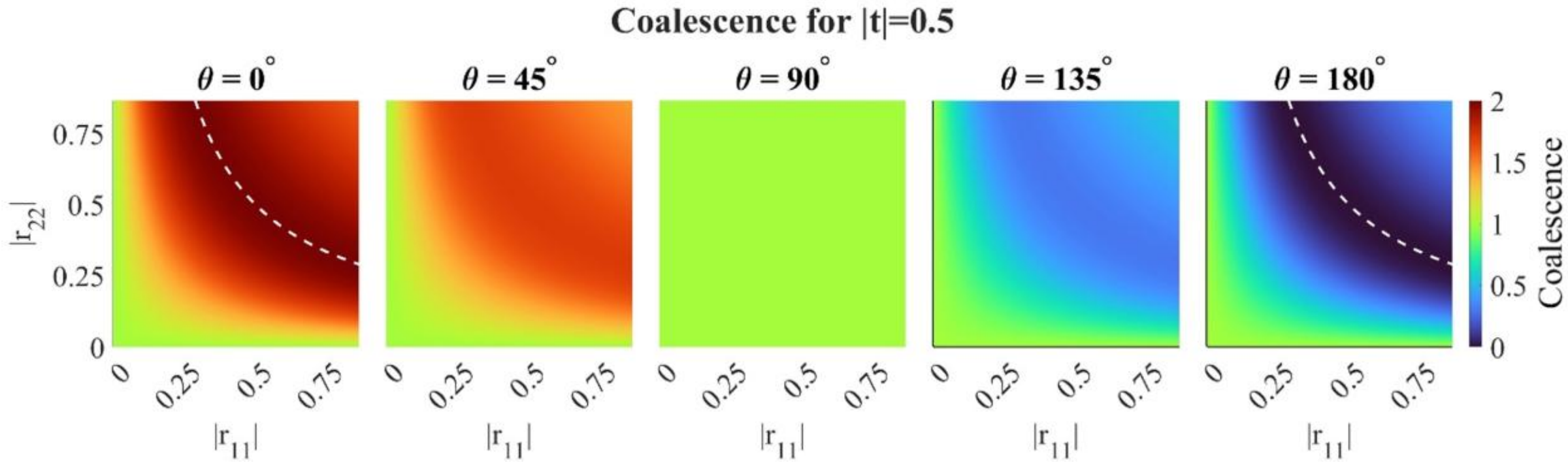

**Fig. 2: Quantum coalescence metric.**

Computed coalescence (C) metric for various $\theta_{rt}, |r_{11}|, |r_{22}|$ values when the transmission coefficient has a fixed amplitude equal to $|t|=0.5$.

### EOT Corrugated Plasmonic Grating

Extraordinary optical transmission (EOT) is an energy tunneling process driven by the periodic arrangement of subwavelength plasmonic apertures that induce momentum on incident free space photons to allow them to couple to highly confined surface plasmon polaritons along the periodic aperture surface [21,22]. This resonant transmission response occurs when the incident radiation wavelength exceeds the fundamental cutoff wavelength of the individual apertures, allowing these surface modes to coherently re-illuminate the apertures through excited electric fields in the formed nanohole or nanoslit gap plasmons leading to funneled light to the far-field with high efficiency [21]. The principles of EOT serves as the foundational mechanism for a wide array of highly compact plasmonic metamaterials [23,24], tunable color filters [25], plasmonic gratings [26,27], and ultra-sensitive biomedical sensing platforms [28] that overcome the limitations of traditional, bulky refractive optics. Here, we use EOT to achieve the amplitude matching conditions required to realize optical quantum interference behavior.

Towards this goal, a plasmonic grating operating in transmission mode is designed. The periodic EOT corrugated grating is embedded in a silicon oxide ($SiO_2$) substrate with periodicity ($p$) equal to 250 nm as seen in Fig. 3a. The grating is composed of a periodic arrangement of 20 nm thick ($t$) silver (Ag) grating with 150 nm width ($b$) forming nanoslits, providing the necessary momentum matching to couple incident photons to surface plasmon polaritons (SPPs). Note that the realistic lossy complex permittivity of metal is used in all our modeling efforts, as explained in more detail in the Methods section. To demonstrate the grating's EOT response, Fig. 3b plots the induced local electric field enhancement in the y-direction demonstrating the formation of gap plasmon dipoles that allow the incident electromagnetic wave to funnel and propagate through the formed nanoslits. The effect is numerically computed along a broad frequency range in Figs. 3b and 3c which shows the transmission of the system dropping in higher frequencies in direct correlation with the decrease in local electric field enhancement. Interestingly, the reflection and transmission coefficient amplitudes become equal at around 180 THz. In Fig. 3d, we demonstrate that this EOT-based plasmonic grating structure provides complete and relatively broadband bosonic coalescence response (*C=0*) near 180 THz due to natively fulfilling the phase shift condition combined with equal transmission and reflection amplitudes. Hence, the presented non-unitary plasmonic grating is designed to mimic a unitary beamsplitter and exhibits bosonic HOM effect at a specific frequency.

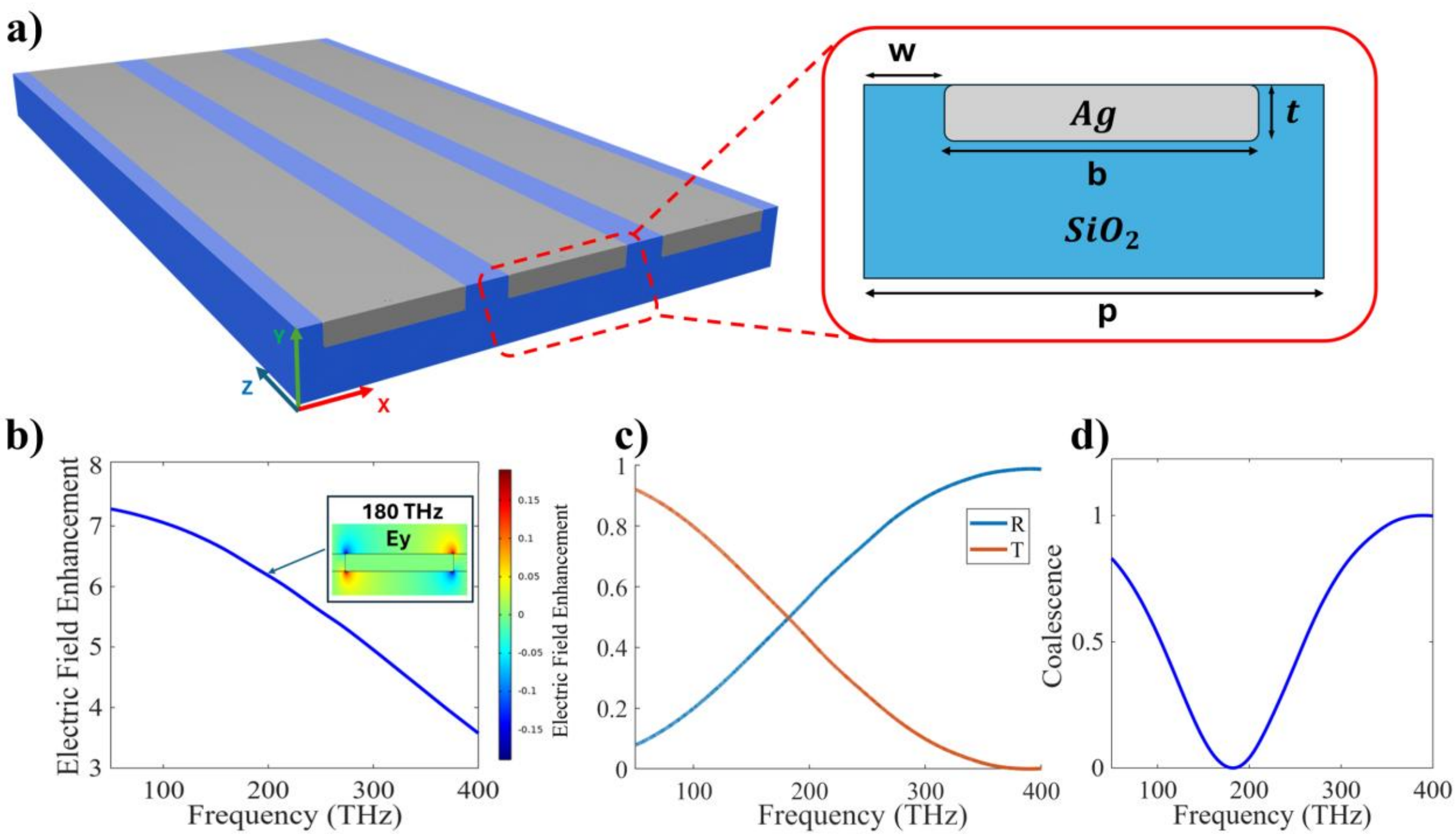

**Fig. 3: EOT corrugated grating.**

(a) Schematic of the EOT plasmonic grating designed to mimic a unitary beamsplitter which exhibits bosonic HOM effect at a specific frequency. (b) Computed electric field enhancement spectrum of the grating with the gap plasmon formation showcased in the inset field enhancement plot computed at the frequency of lowest coalescence. (c) Computed spectrum of reflectance and transmittance. (d) Calculated quantum coalescence response spectrum of the plasmonic grating.

The EOT response operational frequency can be engineered by varying the periodic nanoslit width defined as $w=p-b$ in Fig. 3a. All other grating dimensions are kept fixed to values used to derive Fig. 3 results. By changing the dielectric gap size, the momentum matching condition of the periodic grating is varied and thus the permissible incident radiation EOT coupling frequencies. This directly alters the frequency point where the transmission and reflection amplitudes intersect which, subsequently, changes the grating's quantum coalescence response, as depicted in Fig. 4a. Alternatively, the thickness of the silver grating can be adjusted, while all other grating dimensions are kept equal to values used to derive Fig. 3 results, leading to different frequency ranges where the plasmonic grating exhibits the strongest coalescence response. Such varying bosonic coalescence results cannot occur with an ordinary unitary beamsplitter device.

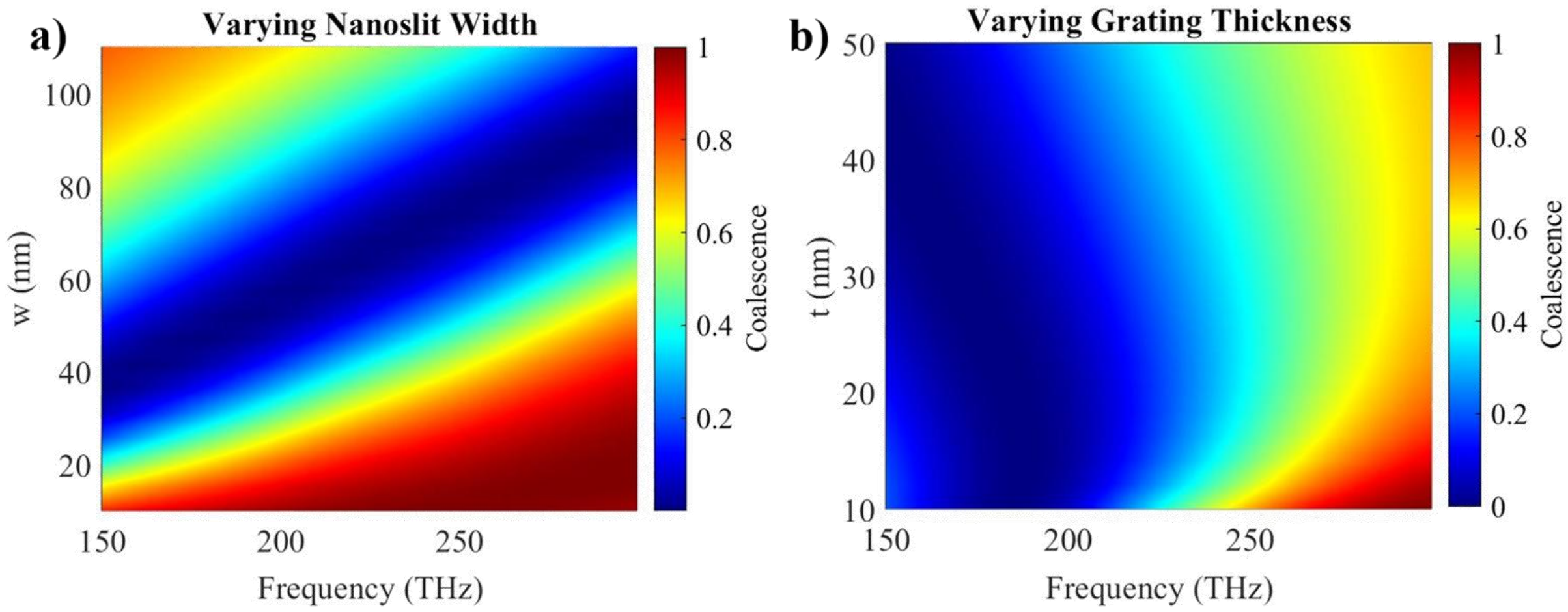

**Fig. 4: EOT corrugated grating coalescence results varying with geometry.**
(a)-(b) Contour plots of plasmonic grating's quantum coalescence performance plotted over varying (a) periodic nanoslit width ($w$) and (b) grating thickness ($t$) while all other dimensions are the same as the ones used to obtain Fig. 3 results.

### Quantum Plasmonic Metasurface

The previously presented plasmonic grating cannot realize fermionic anti-bunching quantum interference response where coalescence becomes $C=2$. To achieve this non-unitary beamsplitter functionality, a secondary shorter width nanostripe resonator array needs to be placed atop the grating, realizing a composite plasmonic metasurface design [29] with schematic depicted in Fig. 5a. The schematic in this figure consists of a 2.5 nm titanium dioxide ($TiO_2$) gap layer (*g*) chosen for its high relative permittivity value which is equal to 5.76 and a periodic array of 50 nm wide (*r*) and 20 nm thick (*h*) Ag nanostripes placed on top. This intricate geometry induces a circulating electric field within the formed nanogap, which due to Ampère's law, drives a strong perpendicular out-of-plane magnetic field resonance as seen in the insets of Fig. 5b where the electric and magnetic fields at the resonance are plotted. In the same figure, the electric field enhancement frequency spectrum is demonstrated when computed at the formed $TiO_2$ nanogap. Interestingly, the highest electric enhancement is obtained around 201 THz indicating a resonant response. The induced magnetic currents at this resonance reradiate to form a sharp Fano response when interfering with the EOT grating reflection, as depicted in Fig. 5c where the reflection and transmission coefficients are presented. The realized abrupt phase shift due to the magnetic nanogap resonance causes the relative phase difference of the reflected and transmitted fields to become equal to zero close to 201 THz, which is at the resonance as seen in Fig. 5d. The different periodic upper nanostripe and lower formed nanoslit width dimensions also create a pronounced asymmetric reflection response while keeping transmission coefficients equal due to reciprocity ($|\boldsymbol{t_{12}}| = |\boldsymbol{t_{21}}| = |\boldsymbol{t}|$), as demonstrated in Fig. 5c, which allows multiple frequencies to fulfill the amplitude matching requirements realizing optical quantum interference leading to quantum coalescence and anti-coalescence.

Hence, the metasurface approximately achieves the amplitude ($|r_{11}| = 0.42, |r_{22}| = 0.69, |t| = 0.42$) and phase ($\theta_{rt} = -2.4^{\circ}$) quantum interference matching conditions in a narrow frequency band at 201 THz leading to anti-coalescence (*C=2*) plotted in Fig. 5e. Conversely, since the destructive interference condition for coalescence only needs the naturally occurring 180-degree phase difference, the metasurface achieves a wider frequency range of bosonic coalescence (*C=0*), minimized at 213 THz due to the following quantum interference matching conditions: $|r_{11}| = 0.68, |r_{22}| = 0.70, |t| = 0.68, \theta_{rt} = 174^{\circ}$ . This demonstrates an almost perfect contrast optical quantum interference transition that is impossible to achieve by other relevant nanophotonic structures (see Supplementary Information document section 2 and its Table S1 for a direct comparison with previous related works). Moreover, unlike other designs that require substantial fabrication modifications, physical manipulation or environmental stimulation, the currently presented composite nanostructure achieves efficient dual state quantum interference simply by slightly switching the frequency of the incident single photons by varying the laser pump wavelength. Experimental considerations relevant to coalescence measurement bandwidth and tunable single photon sources that can be used to measure the presented quantum optical interference effects are presented in Supplementary Information sections 3 and 4, respectively. Finally, it should be noted that non-unitary beam splitters, like our currently proposed design, are lossy optical devices whose scattering matrices do not satisfy unitarity, meaning total output intensity is less than input intensity mainly due to absorption losses. The absorption of our proposed metasurface design is asymmetric due to its geometry and can decrease photon survival probability in an asymmetric way. The asymmetric absorption is plotted in Fig. S1 in the Supplementary Information section 5 along with several comments relevant to its effect on realistic measurements.

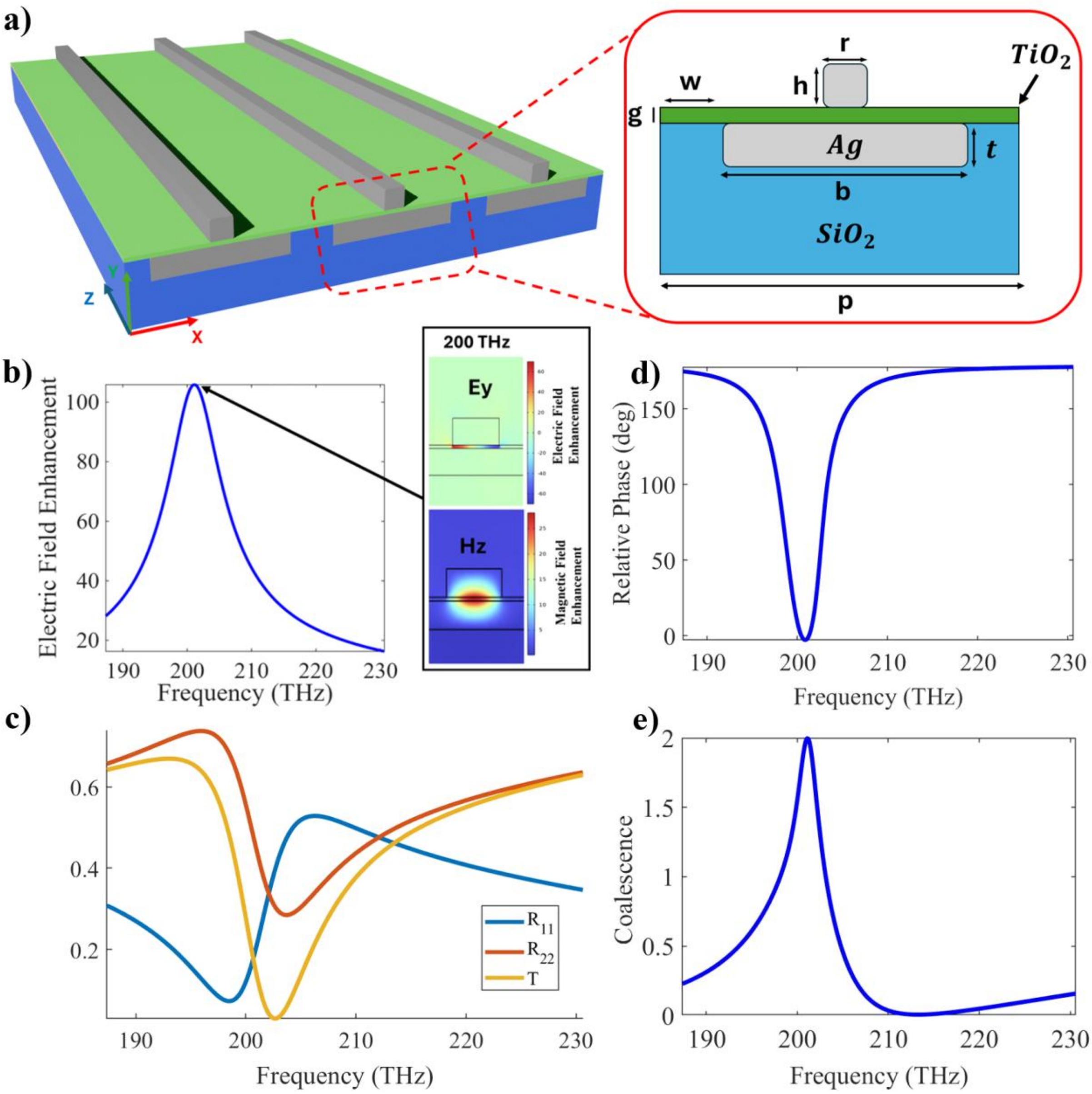

**Fig. 5: Quantum plasmonic metasurface results.**

(a) Schematic of the quantum plasmonic metasurface that exhibits controllable optical quantum interference characterized by varying coalescence. (b) Computed electric field enhancement spectrum of the metasurface, where the circulating electric field and corresponding magnetic field formations are depicted in the insets demonstrated at the highest field enhancement resonant frequency. (c) Calculated reflectance, transmittance, and absorption. (d) Spectra of the transmission and reflection coefficients phase difference. (e) Quantum metasurface computed coalescence values spectra.

The operational frequency and bandwidth of the optical quantum interference can be engineered by varying specific geometric parameters of the realized composite plasmonic metasurface. Modifying the dimensions of the lower corrugated grating primarily dictates the EOT momentum matching condition, as was demonstrated in the previous section. Adjusting the EOT grating's formed periodic nanoslit width ($w$) shifts the allowable SPP coupling frequencies, directly altering where the transmission and reflection amplitudes intersect to allow strong bosonic coalescence characterized by *C=0* with results shown in Fig. 6a. Furthermore, changing the lower plasmonic grating thickness ($t$) tunes the EOT response leading to different reflection and transmission coefficient amplitude-matching conditions required to realize both coalescence and anti-coalescence as seen in Fig. 6b. The anti-coalescence fermionic response characterized by *C=2* is governed by the upper shorter width nanostripe's localized gap-plasmon resonance since it does not exist in the case of bare EOT grating. By altering the width of the top Ag nanostripe ($r$), the resonant condition of the realized nanocavity shifts, allowing for precise control over the onset frequency of the magnetic resonance and subsequent resulted phase shift, leading to a controllable anti-coalesce response as demonstrated in Fig. 6c. Interestingly, in the same figure, other resonance modes exist in higher frequencies outside of the currently studied frequency range that also create anti-coalescing effects, but due to their weaker field enhancements they cannot achieve the same level of anti-coalescence amplitudes. Modifying the $TiO_2$ nanogap thickness ($g$) allows for varying bandwidth by altering the cavity's resonant spectrum as seen in Fig. 6d, where the inset depicts the subregion focused on the strong fermionic response characterized by *C=2*. Larger values of nanogap thickness increase the mode volume of the cavity resulting in decreased electric field confinement. This reduction in field diminishes the phase manipulation capability of the composite quantum metasurface, representing a fundamental tradeoff limiting anti-coalescence response (*C=2*) in the case of $g > 4\ nm$.

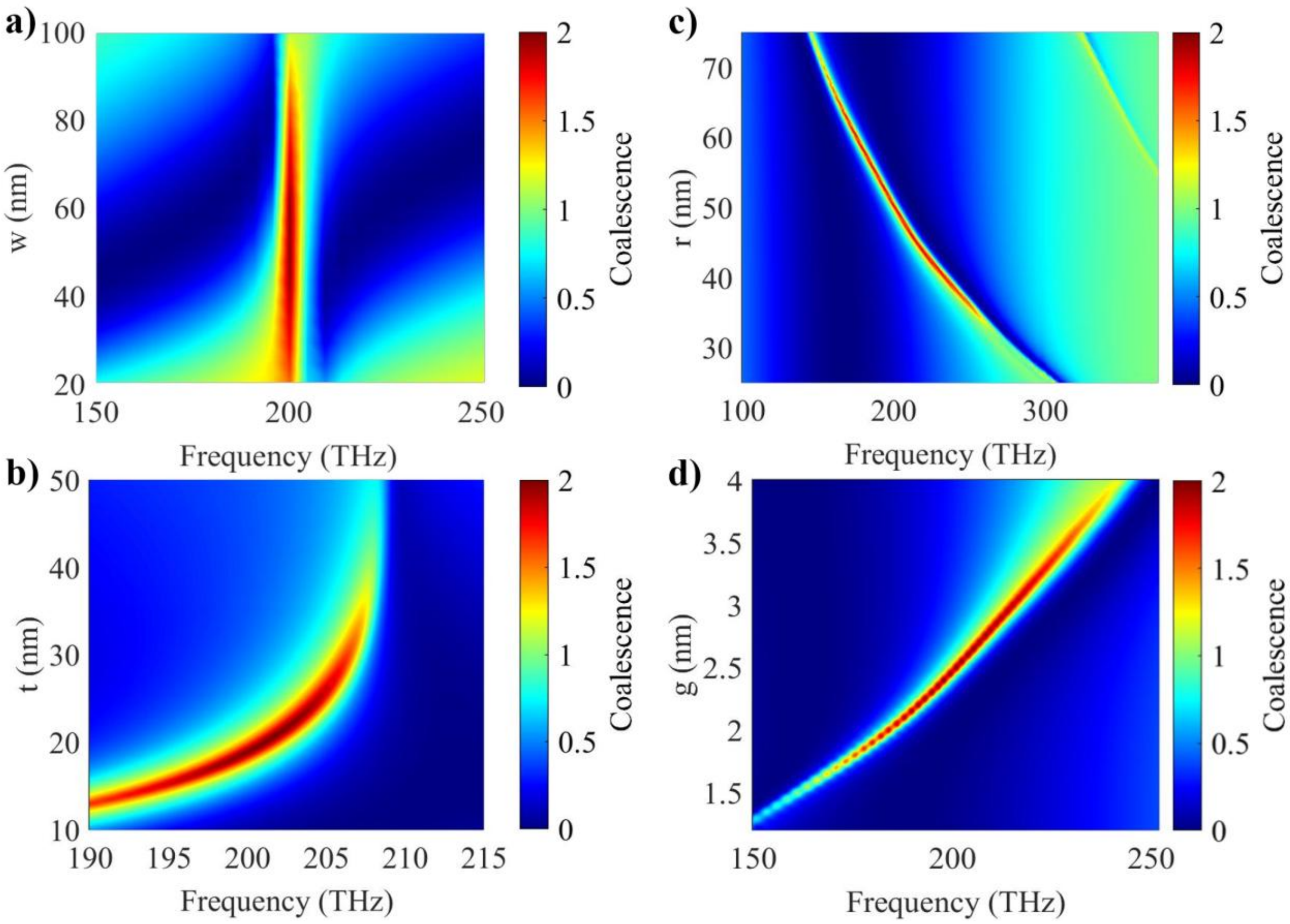

**Fig. 6: Quantum plasmonic metasurface coalescence results varying with geometry.** (a)-(d) Contour plots of composite metasurface quantum coalescence performance plotted over varying (a) nanoslit width ($w$), (b) lower plasmonic grating thickness ($t$), (c) top nanostripe width ($r$), and (d) nanogap thickness ($g$) while all other dimensions are the same as the ones used to obtain Fig. 5 results.

It is worth noting that traditional optical systems are usually comprised of free space arrangements of optical devices distributed across stabilized optical tables. However, modern advancements have allowed the miniaturization of these complex architectures into photonic integrated circuits. These have demonstrated remarkable success in scaling quantum networks, offering enhanced stability, reduced footprints, and the potential for CMOS compatibility [30,31]. While the currently presented metasurface operates in free-space, here, we also prove that it can also be compatible with such on-chip quantum photonic platforms. The major advantage that the metasurface has to offer in terms of its photonic integration is its ability to perform effectively over many different angles of incidence. As seen in Fig. 7, the metasurface is almost unaffected by the angle of incidence from both ports until an incident angle of 50 degrees, a useful functionality that will allow the metasurface to couple into the highly confined guided modes of a photonic integrated circuit without the need of adiabatic waveguide-based coupling. Since it is also fabricated using CMOS friendly materials, it should be compatible with current integrated photonic circuits and easily couple to the evanescent tail along their guided optical modes [30]. Finally, the fabrication of the presented quantum plasmonic metasurface is feasible since a similar composite metasurface design has been experimentally realized to explore exceptional points in classical optics, i.e., different than the currently studied quantum optical response [32]. Several additional comments regarding the potential fabrication process of the presented metasurface are included in Supplementary section 6. Moreover, quantum nonlocal effects and how they affect the response of the presented metasurface due to the close to 2 nm nanogap have been computed and presented in Supplementary section 7.

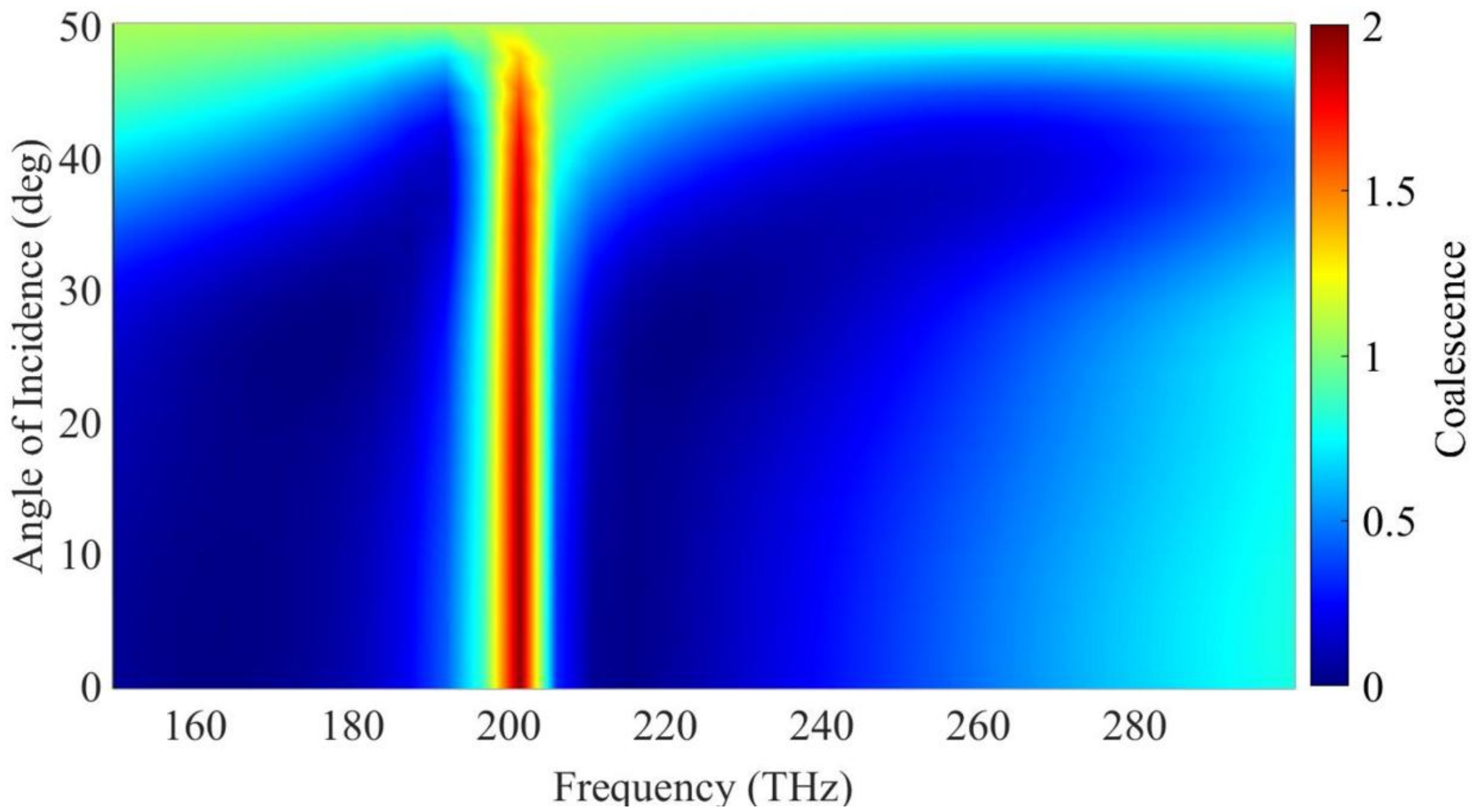

**Fig. 7: Quantum plasmonic metasurface varying incident angle operation.**

Contour plot of metasurface quantum coalescence performance plotted over varying angles of incidence from both ports.

## Discussion

To conclude, in this work we present new asymmetric quantum plasmonic metasurface designs operating in transmission mode that allow for efficient and controllable manipulation over the optical quantum interaction of two incident single photons. More specifically, we demonstrate a composite metasurface design that can work as a non-unitary beamsplitter achieving both bunching and anti-bunching of single photon pairs just by varying its frequency of operation. The presented quantum metasurface design combines a plasmonic grating corrugated by periodic nanoslits that can achieve extraordinary optical transmission with additional nanostripes on top forming nanocavities characterized by a gap plasmon magnetic resonance. The latter resonance leads to the fermionic anti-bunching response in a narrow frequency spectrum in addition to the more broadband bunching response mainly coming from the corrugated grating.

The quantum photon-photon interactions from the presented new quantum metasurface configuration are analyzed based on classical full-wave electromagnetic simulations with results introduced in the four quantum path probability amplitudes that lead to the calculation of the quantum coalescence. This quantum metric accurately portrays the presented bunching and anti-bunching interference responses. It is plotted in different frequencies and for varying metasurface dimensions demonstrating the unique control of photon-photon interactions varying from bosonic to fermionic. The presented efficient optical quantum interference control can be used to implement emerging quantum logic gates that will be utilized in optical quantum computing systems, extremely sensitive quantum optical sensors for high-precision measurements that are robust against environmental noise and improved secure quantum communication networks. The realized ability to artificially invert fundamental particle statistics provides a completely new paradigm for simulating complex, strongly correlated synthetic quantum matter directly on a chip, alongside applications in remote quantum sensing, subwavelength edge detection, and noise-resilient rotation metrology. On a more general note, the findings of our current work provide insights into the transformative aspects that metasurfaces and the field of plasmonics can accomplish in the nascent field of quantum photonic engineering.

## Methods

### Numerical modeling

We use finite element method based full-wave accurate simulations to compute the complex reflection and transmission coefficients of the present metasurfaces. To create a high-fidelity metasurface simulation, the process begins with a meticulous approach to designing the unit cell geometry. Rather than using idealized, sharp-edged blocks, it is essential to define geometries with rounded filleted edges. This prevents unphysically induced field singularities at sharp corners and more accurately mimics the rounded edge profiles resulting from standard nanofabrication techniques like electron-beam lithography. In our simulations our filleted edges had a radius of 2 nm, with all other dimensions of the metasurface being described in each section. Once its geometry is established, the physics environment is organized by creating distinct wave equation nodes for the air, dielectric, and metallic domains.

For non-dispersive lossless dielectric media, the model utilizes constant real refractive index values. However, for metallic lossy components dispersion accuracy is prioritized by manually defining the complex refractive index. More specifically, by importing the real ($n$) and imaginary ($\kappa$) components from tabulated experimental data [33], we define the complex permittivity of metals as $\epsilon_r = (n + i\kappa)^2$. This ensures the simulation captures the correct plasmonic or reflective behavior of the metallic parts of the presented nano configurations. To finalize the design, the metasurface performance is quantified using electromagnetic wave probes. These tools are strategically placed to monitor the field amplitudes, allowing for the direct calculation of reflection and transmission coefficients. By integrating along these probes, we derive the complex S-parameters that provide the phase and amplitude values of reflection and transmission coefficients from both directions as depicted in Fig. 1.

## Data Availability

The data supporting the results reported in this paper are available from the authors upon reasonable request.

## Acknowledgements

This work was partially supported by the Penn State Applied Research Laboratory Walker Fellowship and the National Science Foundation (NSF) DMR-2224456.

## Author contributions

L.Q. and S.S. designed the metasurface and carried out the electromagnetic simulations and numerical calculations. C.A. conceived the idea and supervised the project. All authors analyzed the data and contributed to writing the manuscript.

## Competing Interests

The authors declare no competing interests.

# Supplementary Information

## Controlling Optical Quantum Interference with Plasmonic Metasurfaces

Lawrence M. Qin,[1,2] Sky Semone,[1,2] Matthew J. Brandsema,[2] and Christos Argyropoulos[1,*]

[1]Department of Electrical Engineering, Pennsylvania State University, University Park, PA, 16803 USA

[2]Applied Research Labs, Pennsylvania State University, University Park, PA, 16803 USA

*cfa5361@psu.edu

### 1. Lossy Asymmetric Beamsplitter Quantum Theory

This section establishes the theoretical quantum framework for describing two-photon interference in a lossy and potentially asymmetric non-unitary beamsplitter. Conventional unitary beamsplitter theory is extended to account for optical loss through the inclusion of Langevin noise operators, allowing the output photon-number probabilities to be expressed in terms of the complex transmission and reflection coefficients of the system. These probabilities are then used to quantify two-photon coalescence and anti-coalescence by comparing the quantum interference probability with the corresponding classical probability for distinguishable photons. In particular, the resulting coalescence parameter demonstrates how the relative amplitudes and phases between the reflected and transmitted fields determine the nature of the two-photon interaction, enabling the system to exhibit photon bunching, photon anti-bunching, or classical-like behavior.

To achieve and manipulate quantum two-photon interference, we should consider the conditions that must be set to calculate the joint probability of photons exiting from Port 1 and Port 2, as seen in Fig. 1 of the main text. Standard beamsplitter theory assumes a unitary condition where $|t|^2 + |r|^2 = 1$, where $t, r$ are the transmission and reflection coefficients, respectively. This requires the relationship between the input annihilation operators $(\hat{a}_{in}, \hat{b}_{in})$ and output operators $(\hat{a}_{out}, \hat{b}_{out})$ to strictly rely on transmission $(t)$ and reflection $(r)$ coefficients. However, realistic devices are lossy and dispersive. In this case, we need to loosen the transmission and reflection coefficient restrictions ($|t(\omega)|^2 + |r(\omega)|^2 \leq 1$) and introduce Langevin noise ($F$) operators to account for loss in order to model lossy quantum interactions [1]. These interactions are also going to be asymmetric, thus forward ($t, r_{11}$) and backward ($t, r_{22}$) propagation coefficients must be considered separately and are connected to their respective annihilation operators. These relationships are written in Eq. S1 [1]:

$$\begin{aligned} \hat{a}_{out} &= \hat{a}_{in}(r_{11} + t) + F_a, \\ \hat{b}_{out} &= \hat{b}_{in}(r_{22} + t) + F_b. \end{aligned} \quad (S1)$$

To ensure that the commutation relations between operators remain valid in the presence of the introduced loss, the input channels for both propagations and their corresponding absorption must also be independent as shown in the equations below [1]:

$$\begin{aligned} [\hat{a}_{in} \quad F_a] = [\hat{a}_{in} \quad F_a^\dagger] = [\hat{a}_{in} \quad F_b] = [\hat{a}_{in} \quad F_b^\dagger], \\ [\hat{b}_{in} \quad F_b] = [\hat{b}_{in} \quad F_b^\dagger] = [\hat{b}_{in} \quad F_a] = [\hat{b}_{in} \quad F_a^\dagger]. \end{aligned} \quad (S3)$$

We can then put the two photons incident together to form the state vector $|\psi\rangle$ and normalize it by normalizing the amplitude $\psi$ as seen in Eq. S4 [1]:

$$\begin{aligned} |\psi\rangle &= \int_0^\infty d\omega\psi(\omega)\, \hat{a}_{in}^\dagger(\omega)\hat{b}_{in}^\dagger(\omega)|0\rangle, \\ \int_0^\infty d\omega \int_0^\infty |\psi(\omega)|^2 &= 1. \end{aligned} \quad (S4)$$

The state vector can be represented as expectation values ($\widehat{N}$) of the number operators defined for each output port [1]:

$$\begin{aligned} \widehat{N}_a(\omega) &= \int_0^\infty d\omega\hat{a}_{out}^\dagger(\omega)\, \hat{a}_{out}(\omega), \\ \widehat{N}_b(\omega) &= \int_0^\infty d\omega\hat{b}_{out}^\dagger(\omega)\, \hat{b}_{out}(\omega). \end{aligned} \quad (S5)$$

We can solve both expectation values by applying the output operators for each port ($\hat{a}_{out}(\omega), \hat{b}_{out}(\omega)$) to our state vector ($|\psi\rangle$) over all frequencies. To get our answer in terms of reflection and transmission responses, we convert the amplitude of each output port in terms of their reflection and transmission coefficients. The calculation for the first expectation value is given below in Eq. S6 [2]:

$$\begin{gathered}\langle \hat{N}_a \rangle = \int_0^\infty d\omega \, \langle \psi | \hat{a}_{out}^\dagger(\omega) \hat{a}_{out}(\omega) | \psi \rangle \,, \\ \hat{a}_{out}(\omega)|\psi\rangle = t(\omega) r_{11}(\omega) \int_0^\infty d\omega \psi(\omega) \hat{a}_{in}(\omega)|0\rangle = \\ = 4 \int_0^\infty d\omega t(\omega) r_{11}(\omega) \int_0^\infty d\omega |\psi(\omega)|^2 \,, \\ \langle \hat{N}_a \rangle = |t + r_{11}|^2. \end{gathered} \tag{S6}$$

The joint probability ($P$) of detecting these photons ($n$) with a high quantum efficiency ($\eta$) and a relatively large coincidence counting window ($T$) can then be calculated using the Kelley Kleiner counting formalism given in Eq. S7 [2]:

$$P(n, T) = \left\langle \frac{\left(n\hat{I}\right)^n}{n!} e^{-\eta t \hat{I}} \right\rangle. \tag{$S7$}$$

This allows us to find the mean photon number for each port based on the expectation values of the number operators which gives us the following probabilities [2]:

$$\begin{gathered} P(2,0) = \frac{1}{2} \langle N_a (N_a - 1) \rangle, \\ P(0,2) = \frac{1}{2} \langle N_b (N_b - 1) \rangle, \\ P(1,1) = \langle N_a N_b \rangle. \end{gathered} \tag{$S8$}$$

Finally, we can apply the solved expectation values calculated in Eq. S6 to get the quantum probability of each state as seen in Eq. S9 with $\theta_{rt}$ representing the phase difference between the reflected and transmitted waves which are identical formulas to Equations (1)-(3) in the main paper [2]:

$$P(2,0)_q = 2|t|^2|r_{11}|^2,$$
$$P(0,2)_q = 2|t|^2|r_{22}|^2, \qquad (S9)$$
$$P(1,1)_q = |t|^4 + |r_{11}|^2|\, r_{22}|^2 + 2|t|^2|r_{11}||r_{22}|\cos(\theta_{rt}).$$

## 2. Comparison of Coalescence Values in Recent Relevant Research

To place the performance of the proposed metasurface in context with relevant literature, Table S1 compares the coalescence values reported in previous studies of varying two-photon interference using metasurfaces [3–5]. These studies demonstrate the ability to manipulate Hong–Ou–Mandel interference and control the transition between photon coalescence and anti-coalescence through engineered amplitude and phase responses. The reported ranges provide a useful benchmark for evaluating the performance of the present approach. In comparison, the metasurface demonstrated in our current work achieves a broader range of coalescence values, enabling more extensive control over the transition between photon bunching and anti-bunching behaviors.

**Table S1**

| Reference | Photonic Design | Coalescence Range |
|---|---|---|
| **[3]** | Unitary metasurface (Theory) | 0.01 - 1.79 |
| **[4]** | Non-unitary metasurface (Experiment) | 0.35 - 1.71 |
| **[5]** | Phase-change metasurface (Theory) | 0.023 - 1.755 |
| **This Study** | Non-unitary plasmonic metasurface (Theory) | 0.001 - 2.000 |

## 3. Experimental Considerations of Coalescence Measurement Bandwidth

The principal effect of near-perfect coalescence and anti-coalescence at nearby frequencies has potential applications where tunable coalescence is desired by changing the frequency of incident light. Because the metasurface beamsplitter in this study shows coalescence between pairs of photons, we assume that a potential experimental configuration would consist entirely of degenerate photon pairs generated through spontaneous parametric down-conversion (SPDC), which is the most widely used method to generate single photon pairs. In this section, we highlight existing technologies that can tune the photon-pair generation over a substantial frequency range and also exhibit the required narrow spectral linewidth to experimentally realize the presented near-complete coalescence and anti-coalescence results.

The full width at half maximum (FWHM) value of the anti-coalescence in Fig. 5e of the main paper has an approximate value of 2 THz or 2000 GHz, using a coalescence value of 1.5 to set bounds on the peak at 200 THz and 202 THz, respectively. In terms of wavelengths, this is a reasonably broad range of 1484-1499 nm that needs to be considered when selecting a source for SPDC photon pairs.

The photon bandwidth is an experimentally controllable parameter, and it should not be larger than the metasurface resonance linewidth. For this reason, our device should be paired with an SPDC source that has a bandwidth <15nm, which can be found in Type-II phase matched periodically poled lithium niobate (PPLN) waveguides. More specifically, bright narrowband sources of photon pairs at optical telecommunication wavelengths have been demonstrated based on type-II PPLN waveguide producing photon pairs at the resonance frequency of our metasurface with a measured FWHM bandwidth of approximately 1 nm (about 125 GHz) [6], i.e., much lower values than the metasurface anti-coalescence bandwidth. The same experiment produced approximately 40 nm bandwidth from a type-0 PPLN device, illustrating the substantially narrower frequency response obtained with type-II PPLN. Other telecom type-II PPLN experiments have reported even narrower bandwidths of approximately 0.85 nm at 1540 nm and 0.7 nm at 1550 nm, corresponding to approximately 100 and 90 GHz, respectively [7]. More generally, reported type-II PPLN waveguide sources in the 1540–1560 nm region span approximately bandwidths of 0.7–2.4 nm, while conventional type-0 PPLN sources commonly have bandwidths of tens of nanometers [8].

Besides Type-II phase matching, cavity-enhanced SPDC provides an alternative and more direct route to making the photon bandwidth much narrower than the currently presented resonance linewidth. In conventional single-pass PPLN waveguide SPDC sources, the phase-matching bandwidth is typically relatively broad; for example, a PPLN crystal has been measured to produce a signal bandwidth of approximately 1 THz, which is still lower than the anti-coalescence bandwidth of our presented metasurface. By placing the PPLN crystal inside a resonant cavity, the same SPDC process can instead be restricted to narrow cavity modes, reducing the photon linewidth to the tens-of-MHz regime [9,10]. An experimental implementation using a 6-mm PPLN crystal reported cavity modes with approximately 48 MHz linewidth and a free spectral range of approximately 4.5 GHz [11].

More generally, cavity-enhanced SPDC sources have demonstrated linewidths ranging from several MHz down to the sub-MHz regime through appropriate choices of cavity length and finesse. Reported examples include 6.6 MHz for a short high-finesse cavity, approximately 2 MHz for a longer cavity, and 0.66 MHz for an approximately 2.5 m effective cavity length [9]. Directly at telecom wavelengths, a triple-resonant SPDC source based on a periodically poled potassium titanyl phosphate (PPKTP) nonlinear crystal has produced photon pairs near 1560 nm with an 8 MHz bandwidth [10]. Although this latter demonstration used PPKTP rather than PPLN, it establishes that MHz-scale cavity-SPDC linewidths are experimentally achievable in the telecom band. Such bandwidth values are substantially lower compared to our quantum metasurface anti-coalescence bandwidth of approximately 2 THz = $2 \times 10^6$ MHz.

## 4. Experimental Considerations of Tunable Single Photon Sources

The frequency separation of approximately 12 THz between the coalescence and anti-coalescence responses presented in the main paper should be considered carefully when discussing experimental implementation. For SPDC sources, the generated photon pair wavelength is determined by the phase-matching conditions of the nonlinear medium and can be tuned through several experimentally accessible parameters, including the pump wavelength, crystal temperature, crystal orientation, and, in periodically poled materials, the quasi-phase-matching period. In particular, periodically poled nonlinear crystals such as PPKTP and PPLN provide practical approaches for wavelength-tunable photon-pair generation. More specifically, highly pure SPDC sources based on PPKTP realizing broad photon wavelength tunability from 1460 to 1675 nm depending on the incident laser wavelength have been experimentally demonstrated in [12]. This range substantially surpasses the wavelength range relevant to the present work, since 13 THz is approximately 80 nm wavelength range. Another similarly broadband SPDC source based on magnesium oxide (MgO):PPLN has been reported, in which the emission wavelength can be controlled by varying the crystal temperature and pump wavelength [13]. These demonstrations indicate that the required spectral region is accessible with appropriately engineered SPDC sources, although rapid switching between 201 and 213 THz would require a laser pump that can also be swept in a fast way between various frequencies.

A further possibility is provided by ultrathin lithium niobate (LN), where the conventional bulk crystal phase-matching restrictions can be substantially relaxed. In sufficiently thin LN films, the interaction length becomes much shorter than in conventional bulk or periodically poled crystals, and momentum conservation in the longitudinal direction becomes less restrictive. Non phase-matched SPDC has been experimentally demonstrated in ultrathin LN films, including a 300 nm-thick LN film [14]. In this regime, the reduced interaction length broadens the phase-matching response and allows photon-pair generation over a substantially wider spectral range without requiring conventional bulk-crystal phase matching. This is particularly attractive for the present application because a broadband SPDC source based on ultrathin LN could potentially provide access to both 1408 and 1499 nm.

The aforementioned tunable single photon sources require a pump laser that can be tuned continuously from approximately 402 THz to 416 THz (703 nm to 746 nm, respectively) which can be realized by using a broadband optical source followed by fast wavelength selection. In particular, supercontinuum light sources can provide broadband emission extending across the visible and near-infrared, while an acousto-optic tunable filter (AOTF) can select narrow spectral channels electronically by changing the applied radio-frequency drive. Commercial and experimental supercontinuum/AOTF systems have demonstrated wavelength selection over broad spectral ranges, including the 1.4–1.5 μm wavelength region relevant to the present device [15,16]. Such architecture could therefore provide convenient access to both operating wavelengths without requiring the nonlinear crystal itself to be tuned over the full 13 THz interval in micro-seconds, without any slow mechanical adjustment.

While this section provides a path to continuous coalescence tuning using technologies readily available at the time of writing, there are many other possible applications and experimental configurations made possible by our presented quantum metasurface. Of note is the possibility of using two SPDC sources emitting single photon pairs at both 201 and 213 THz, respectively, to realize both coalescence and anti-coalescence. By changing the power of each pump laser, the full range of coalescence states provided by the metasurface can be accessed.

## 5. Absorption of the Asymmetric Metasurface

Non-unitary beam splitters, like our currently proposed metasurface design, are lossy optical devices whose scattering matrices do not satisfy unitarity, meaning total output intensity is less than input mainly due to absorption losses. Hence, a complete characterization of the coalescence metasurface requires consideration of its transmittance, reflectance, and absorption spectra over the full operating frequency range. While the beam-splitting functionality is primarily determined by the relative distribution of optical power between the transmitted output channels, the absorbed power provides important information about the underlying resonant response, energy redistribution within the metasurface, absolute photon transmission, and photon survival probability. In particular, the spectral region associated with the designed anti-coalescence coincides with strong field confinement and absorption, which reduces the overall transmission. For practical applications, it is desirable for the absorption to be as small as possible, since use of the device requires retrieving the transmitted and reflected light in the most efficient way possible.

The plot of the reflectance, transmittance, and absorption over the span of the anti-coalescence/coalescence transition is shown in Fig. S1, highlighting with black vertical solid lines the two operating points corresponding to C=2 (anti-coalescence) and C=0 (coalescence) that occur at the respective maximum and minimum C values plotted in the main paper Fig. 5e. Specifically, the reflectance and transmittance are $R_{11} = T = 0.177$ at anti-coalescence which are approximately half values compared to the almost lossless $R_{11} = T = 0.464$ values obtained at coalescence, where the transmittance $T$ is identical for both directions of illumination $T = T_{12} = T_{21}$ since the metasurface is reciprocal structure. Interestingly, when the current non-unitary asymmetric metasurface is illuminated from the opposite side (from Port 2 to 1) the absorption at the anti-coalescence is substantially less due to the increased reflectance $R_{22}$ at this frequency point, as seen in Fig. S1. Hence, we can achieve strong anti-coalescence response but with asymmetric absorption, a unique feature of current metasurface design, leading to different survival probabilities of the transmitted and reflected photons. This is an additional degree of freedom that is possible only with the current asymmetric nanophotonic design and proves that anti-coalescence arises mainly from two-photon interference.

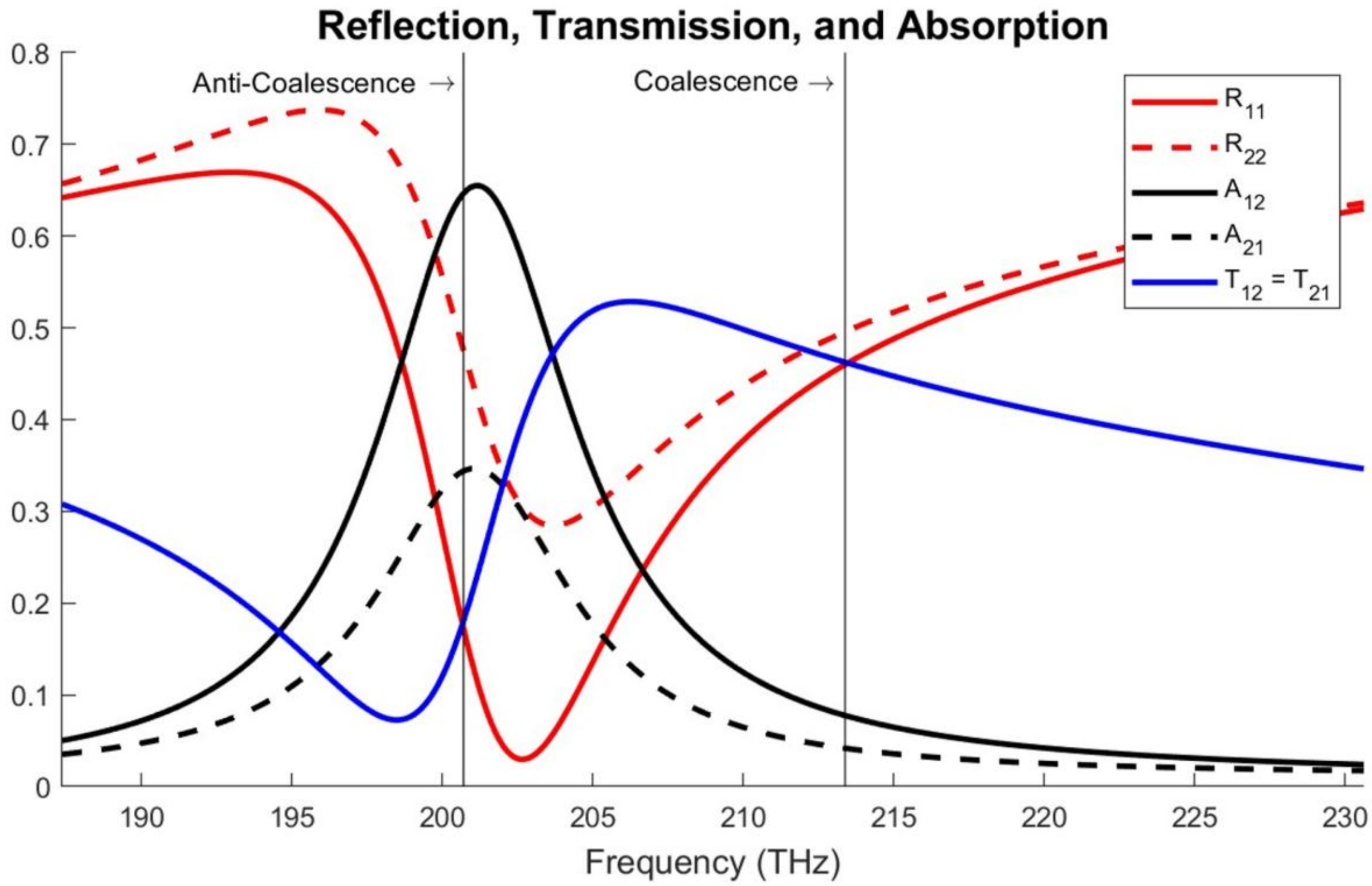

**Fig. S1:** Reflectance, transmittance, and absorption plotted around the frequencies corresponding to anti-coalescence and coalescence of the presented quantum asymmetric metasurface design.

## 6. Potential Metasurface Fabrication Process

The fabrication of the presented quantum composite metasurface is feasible since recently a similar metasurface design has been experimentally realized in [17] to explore exceptional points in classical optics, i.e., different than the currently studied quantum optical response. More specifically, the nanofabrication of the presented asymmetric quantum plasmonic metasurface can begin by etching high anisotropy trenches into a fused silicon dioxide ($SiO_2$) substrate using Inductively Coupled Plasma Reactive Ion Etching (ICP-RIE) or Atomic Layer Etching (ALE) as direct etching of silver is significantly more difficult. Patterning should be achieved through Electron Beam Lithography (EBL), with powers up to 100kV recommended to minimize forward scattering. Along with it, a bilayer resist stack of Methyl methacrylate (MMA) and Poly(methyl methacrylate) (PMMA) should be utilized to create a controlled undercut profile, which prevents metal buildup on sidewalls during deposition. To refine feature fidelity, a cold development around $5^{\circ}C$ is encouraged to slow down EBL reaction with the resist, allowing higher PMMA contrast and ensuring the complete removal of resist residue from the trench floors.

The ingrained silver (Ag) nanoslits can then be deposited via electron beam evaporation under high vacuum to ensure the high directionality required for effective lift-off while reducing potential defects. The substrate should then be planarized through Chemical Mechanical Polishing (CMP) using hard mask stop layers to ensure the Ag is perfectly level with the $SiO_2$ substrate. Following planarization, a 2.5 nm titanium dioxide $TiO_2$ gap layer will be deposited using Atomic Layer Deposition (ALD) for maximum precision and uniformity (i.e., achieving minimum surface roughness), followed by a secondary EBL and evaporation cycle to deposit the top Ag resonators. Hence, it can be concluded that the currently presented metasurface is possible to be fabricated with limited roughness and imperfections and this would be subject of our future work.

Recently, we successfully fabricated and tested a plasmonic metasurface of similar scale using gold nanostripes separated by a 4 nm $TiO_2$ layer [18]. In this previous design, the realized nanophotonic metasurface created field enhancement at a particular frequency in the near infrared, like the beamsplitter model presented in this work, but operating as a plasmonic absorber. The experimental measurements and simulation model showed nearly identical results in reflectance and subsequent nonlinear (i.e., not quantum, different than current work) effects created by the realized metasurface. Hence, the effects of surface roughness were limited and ignored in the simulations without affecting the matching between theory and experimental results.

Finally, it is worth mentioning that there are inherent manufacturing tolerances to all processes that can shift the response of the metasurface or degrade the quality, but the design used in this study allows for sufficient flexibility since geometric features can be fine-tuned to compensate for the fabrication tolerances. An example of such fine-tuning of metasurface parameters is shown in the following section where the coalescence response is plotted when nonlocal quantum metasurface simulations are performed. If variations in the fabricated device geometry are present, they may slightly degrade the quality of the metasurface or detune the resonance but not substantially affect the presented coalescence effect. Once the metasurface is fabricated and tested, the simulation can be revised to incorporate these tolerances and roughness effects and re-optimized to create a final, higher-quality device.

## 7. Nonlocal Metasurface Simulations

In the primary text of the paper, nonlocal effects due to the close proximity of the metallic features were not considered, because previous studies have shown them to be relatively weak for larger than 2 nm gaps [19,20]. Nonlocal quantum effects are expected to lead to a spectral "blueshift" in the metasurface resonance and a slight decrease in the nanogap field enhancement [19]. In our main metasurface geometry, the nanogap is set to 2.5 nm which is expected to also be affected by nonlocal effects. In the following, we perform nonlocal full-wave simulations and explicitly demonstrate that nonlocal effects slightly blueshift the coalescence resonance frequency. However, we also present that by slightly varying the geometrical parameters of the nonlocal metasurface design we can compensate for these effects and restore the coalescence response at the resonance frequency of the presented in the main paper local response.

The nonlocal response of the plasmonic metasurface is modeled by using the hydrodynamic Drude model [20] for metals. The model extends the conventional local Drude model by treating the conduction-electron gas as a compressible fluid. This introduces a spatial-dispersion term into the metallic response, allowing the electron density to vary over length scales comparable to the electronic screening length. As a result, the model captures quantum nonlocal effects due to metal interfaces proximity that are absent from the conventional local response approximation.

For a plasmonic metasurface, the model can be realized by solving Maxwell's equations together with the hydrodynamic equation characterizing the free-electron polarization. The total polarization is separated into free-electron and bound-electron contributions, with the free-electron polarization governed by an equation of the form:

$$-\beta^2\nabla\left(\nabla\cdot\vec{j}\,(\vec{r})\right)+(\omega^2+i\omega\gamma)\,\vec{j}\,(\vec{r})=i\epsilon_0\omega_p\omega\,\vec{E}\,(\vec{r}), \qquad (S10)$$

where $\omega_p$ is the plasma frequency, $\gamma$ is the electron damping factor, and $\beta$ is related to the Fermi velocity. In our simulations, silver is always used for the metallic portions of the plasmonic metasurface and the Drude model parameters used in Eq. S10 are $\omega_p = 2\pi\cdot 2175$ THz and $\gamma = 2\pi\cdot 4.35$ THz [21]. The value of $\beta = v_{F-Silver} = 1.39\cdot 10^6\ m/s$ was chosen based on its use in relevant previous works [20,22,23]. The addition of the $\beta^2\nabla\left(\nabla\cdot\vec{j}\,(\vec{r})\right)$ term in the free-electron polarization introduces nonlocality/spatial dispersion into the metal's electromagnetic response. At the metal-dielectric boundary, the boundary condition $\vec{j}\,(\vec{r}) \cdot \hat{n} = 0$, is added to prevent the free-electron current from flowing normally through the metal boundary [20]. The metasurface is then excited with a plane wave, and quantities such as the electric field enhancement, reflection/transmission spectra, and localized surface-plasmon resonances are calculated. This method has been shown successful in modeling plasmonic metasurfaces with metallic features separated by approximately 2 nm gaps in [19].

The local metasurface coalescence response shown in the main text is reproduced in Fig S2 (blue line) to compare with the nonlocal (NL) model with same dimension metasurface (orange line). The slight "blueshift" in frequency and decrease in anti-coalescence from 2.00 to 1.78 is shown by the NL data. By reoptimizing the nonlocal metasurface with the goal to achieve highest anti-coalescence at the original peak of 201 THz, we find that slightly changing the widths of the upper silver nanostripe array ($r$) from 50 nm to 54 nm, and the lower silver grating width ($b$) from 150 nm to 145 nm can recover nearly the same response as the original model without nonlocal contributions (see yellow line in Fig S2). The anti-coalescence amplitude can also be increased and become equal to two with some additional geometrical optimization of the nonlocal metasurface design. Hence, it can be concluded that while the metasurface response can be slightly affected by nonlocal effects due to the small nanogap, the coalescence response can return to the local resonance frequency with some minor changes of the metasurface geometry.

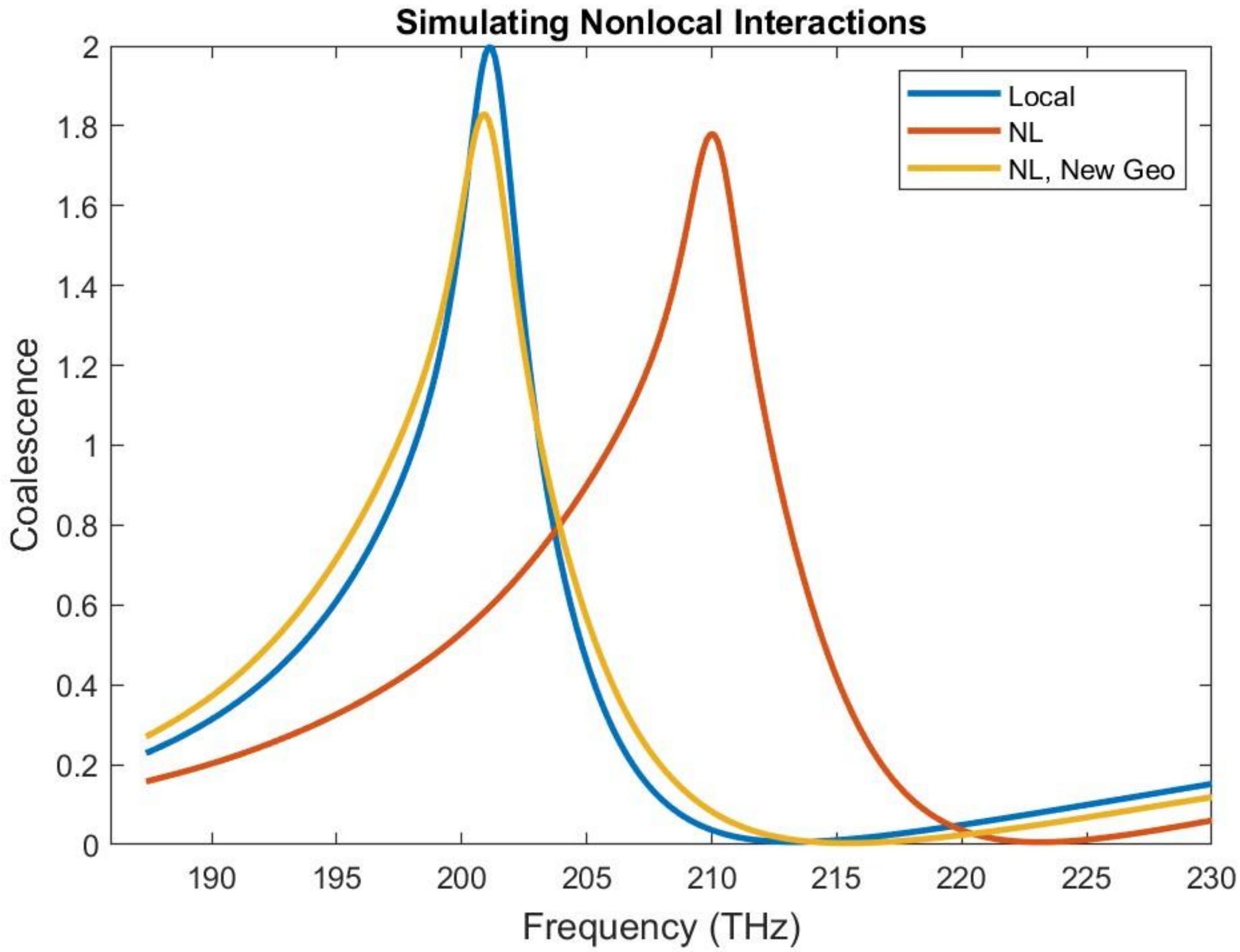

**Fig. S2:** Local metasurface coalescence response (blue line) taken from Fig, 5e in the main text. Applying the nonlocal (NL) model to the same metasurface geometry results in a slight detuning of coalescence frequency (orange line). By fine-tuning the geometry of the nonlocal metasurface, the original target frequency anti-coalescence and coalescence can be obtained (yellow line).